\documentclass[a4paper,11pt]{article}
\usepackage{pos}

\title{Theory Perspective on Heavy Flavour Physics}

\ShortTitle{Theory Perspective on Heavy Flavour Physics}

\author[a,b]{Monika Blanke}
\affiliation[a]{Institute for Astroparticle Physics (IAP), Karlsruhe Institute of Technology, Hermann-von-Helmholtz-Platz 1, D-76344 Eggenstein-Leopoldshafen, Germany}
 \affiliation[b]{Institute for Theoretical Particle Physics (TTP), Karlsruhe Institute of Technology, Engesserstrasse 7, D-76131 Karlsruhe, Germany}
 
       \emailAdd{monika.blanke@kit.edu}

\abstract{
We review the Spring 2022 status of the current $B$ anomalies and their possible interpretation in terms of New Phsyics. We also discuss the discovery potential of targeted LHC and future collider searches for the underlying new particles and their complementarity with low-energy flavour observables.
 \\

{\it Preprint numbers:} TTP22-046, P3H-22-073}

\FullConference{%
  The Tenth Annual Conference on Large Hadron Collider Physics - LHCP2022\\
  16-20 May 2022\\
  online
}

\begin{document}
\maketitle


\section{Introduction}

Before directly producing new heavy particles in high-energy collisions, historically often a first indication for their existence has been the observation of their quantum effects in precision observables. It is hence not surprising that while the LHC experiments ATLAS and CMS have not discovered any new states so far, we have increasing evidence for several anomalies in the flavour and precision sector, namely the so-ccalled $B$ anomalies and the anomalous magnetic moment $(g-2)_\mu$ of the muon. Strikingly, all these observables hint towards the violation of lepton flavour universality.

In this proceedings article, we provide a brief review of the Spring 2022 status of the $B$ anomalies focussing on their potential resolution in terms of New Physics (NP). We also discuss the opportunities to test the NP at the LHC and future particle colliders, and highlight the complementarity between flavour physics observables and direct searches.


\section{\boldmath The $R(D^{(*)})$ anomaly}

A decade has passed since BaBar first announced an anomaly in the data for the lepton flavour universality (LFU) ratios \cite{Lees:2012xj,Lees:2013uzd}
\begin{equation}
R(D^{(*)}) = \frac{\text{BR}(B\to D^{(*)} \tau\nu)}{\text{BR}(B\to D^{(*)} \ell\nu)} \qquad (\ell=e,\mu)\,,
\end{equation}
finding both ratios above their Standard Model (SM) predictions.
Today, having at hand the measurements from BaBar, Belle \cite{Hirose:2016wfn,Hirose:2017dxl,Abdesselam:2019dgh} and LHCb \cite{Aaij:2015yra,Aaij:2017uff,Aaij:2017deq}, the significance of the anomaly amounts to $3.3\sigma$ \cite{HFLAV:2022pwe}. Besides, also the ratio $R(J/\psi)$ is found somewhat above its SM value \cite{Aaij:2017tyk}. On the theory side, the ratios $R(D^{(*)})$ are significantly cleaner than the individual branching ratios, as they are independent of the CKM element $|V_{cb}|$ and the hadronic uncertainties largely cancel in the ratio.

Recently the LHCb collaboration announced a first measurement of \cite{LHCb:2022piu}
\begin{equation}
R(\Lambda_c^+ ) = 0.242 \pm 0.026 \pm 0.040 \pm 0.059\,,
\end{equation}
testing LFU in the related baryonic $\Lambda_b\to \Lambda_c\tau\nu/\ell\nu$ decays. Interestingly this result appears to be a bit below the SM prediction $R(\Lambda_c )_\text{SM} = 0.33 \pm 0.01$ \cite{Detmold:2015aaa},
however the current uncertainties prevent clear-cut conclusions. Notably, however, NP effects  in   $R(\Lambda_c )$ are not independent of the enhancements seen in $R(D^{(*)})$ -- instead, the three ratios are related to each other by a model-independent sum rule \cite{Blanke:2018yud,Blanke:2019qrx}
\begin{equation}
\frac{{R}(\Lambda_c)}{{R}(\Lambda_c)_{\rm SM}}
\simeq 0.262 \frac{{R}(D)}{{R}(D)_{\rm SM}} + 0.738 \frac{{R}(D^*)}{{R}(D^*)_{\rm
      SM}} = 1.15 \pm 0.04\,,
 \end{equation}     
predicting an enhancement of $R(\Lambda_c )$ if the anomaly in $R(D^{(*)})$ holds true. Future more precise determinations of $R(\Lambda_c )$ will hence provide an experimental consistency check of the $R(D^{(*)})$ anomaly.

NP in the underlying $b\to c\tau\nu$ transition\footnote{NP effects in $b\to c e/\mu\nu$ are stringently constrained and cannot account for the observed tension.} can be parametrised by the effective Hamiltonian
\begin{equation}
 {\cal H}_{\rm eff} =  2\sqrt{2} G_{F} V^{}_{cb} \Big[(1+C_{V}^{L}) O_{V}^L +   C_{S}^{R} O_{S}^{R} 
   +C_{S}^{L} O_{S}^L+   C_{T} O_{T}\Big] +\text{h.c.}
\end{equation}
with the dimension-six four-fermion operators
\begin{align*}
  & O_{V}^L  = \left(\bar c\gamma ^{\mu } P_L b\right)  \left(\bar\tau \gamma_{\mu } P_L \nu_{\tau}\right)\,, &&    O_{S}^R  = \left( \bar c P_R b \right) \left( \bar\tau P_L \nu_{\tau}\right)\,,\\
  & O_{T}  = \left( \bar c \sigma^{\mu\nu}P_L  b \right) \left( \bar\tau \sigma_{\mu\nu} P_L \nu_{\tau}\right)\,, &&
   O_{S}^L  = \left( \bar c P_L b \right) \left( \bar\tau P_L \nu_{\tau}\right)  \,.
\end{align*}
The corrensponding Wilson coefficients relate to the possible tree-level NP mediators as follows:
\begin{itemize}
\item
$C_V^L \ne 0$:  This one-dimensional scenario is generated by a heavy $W'$ boson with left-handed couplings \cite{He:2012zp,Greljo:2015mma}. This NP model is put under severe pressure by electroweak precision constraints \cite{Feruglio:2017rjo} and direct LHC searches \cite{Faroughy:2016osc}, and we will not pursue it further.
\item
$C_S^{L,R} \ne 0$: Contributions to the scalar operators are generated by the exchange of a charged Higgs boson $H^-$ \cite{Kalinowski:1990ba,Hou:1992sy,Crivellin:2012ye}. This scenario provides a particularly good fit to the low-energy data, as it is the only one improving the agreement also in the polarisation observable $F_L(D^{(*)})$. A very large branching ratio for the decay $B_C\to\tau\nu$ is predicted in this case, however this cannot be excluded at present \cite{Blanke:2018yud,Blanke:2019qrx,Aebischer:2021ilm}.
\item
$C_V^L, C_S^R \ne 0$: This combination of Wilson coefficients is present in models with an $SU(2)_L$-singlet vector leptoquark $U_1$ \cite{Alonso:2015sja,Calibbi:2015kma,Fajfer:2015ycq,Bordone:2017bld}. 
\item
$C_V^L,C_S^L=-4C_T \ne 0$: The $SU(2)_L$-singlet scalar leptoquark $S_1$ gives rise to this scenario \cite{Deshpande:2012rr,Tanaka:2012nw,Sakaki:2013bfa}.
\item
$C_S^L = 4C_T \ne 0$: The $SU(2)_L$ doublet scalar leptoquark $S_2$ can solve the $R(D^{(*)})$ anomaly provided that its couplings are CP-violating \cite{Becirevic:2018afm}.
\end{itemize}
All of the leptoquark scenarios listed above provide a good fit to the $R(D^{(*)})$ data, see \cite{Blanke:2018yud,Blanke:2019qrx,Murgui:2019czp,Shi:2019gxi,Aebischer:2019mlg}, and they are less stringently constrained by direct LHC searches. Note that we do not consider NP models with light right-handed neutrinos, as they are stringently constrained by LHC data.

A particularly relevant LHC constraint on NP models for the $R(D^{(*)})$ anomaly stems from mono-$\tau$ searches, as crossing symmetry relates $b\to c\tau\nu$ to the process $pp \to X\tau\nu$ \cite{Greljo:2018tzh}. Already with the currently available Run 2 data, stringent constraints are available, and future runs of the (HL-)LHC are expected to test all possible NP scenarios.

In particular, $W'\to \tau\nu$ resonance searches exclude the charged Higgs solution for $m_{H^-} > 400\,\text{GeV}$ \cite{Iguro:2018fni}, while searches for light charged Higgses are plagued by the huge SM $W\to\tau\nu$ background. This problem can be circumvented by requiring an additional $b$-tagged jet in the final state, as shown in \cite{Blanke:2022pjy}. For simplicity, in that study the minimal coupling scenario
\begin{equation}
{\cal L}_\text{int}=
 + y_Q H^- (\overline{b} P_R c)
- y_\tau H^- (\overline{\tau} P_L \nu_{\tau}) +\text{h.c.}
\end{equation}
has been assumed, since additional charged Higgs couplings are not expected to qualitatively alter the conclusions. 
\begin{figure}
\centering
\includegraphics[width=0.45\textwidth]{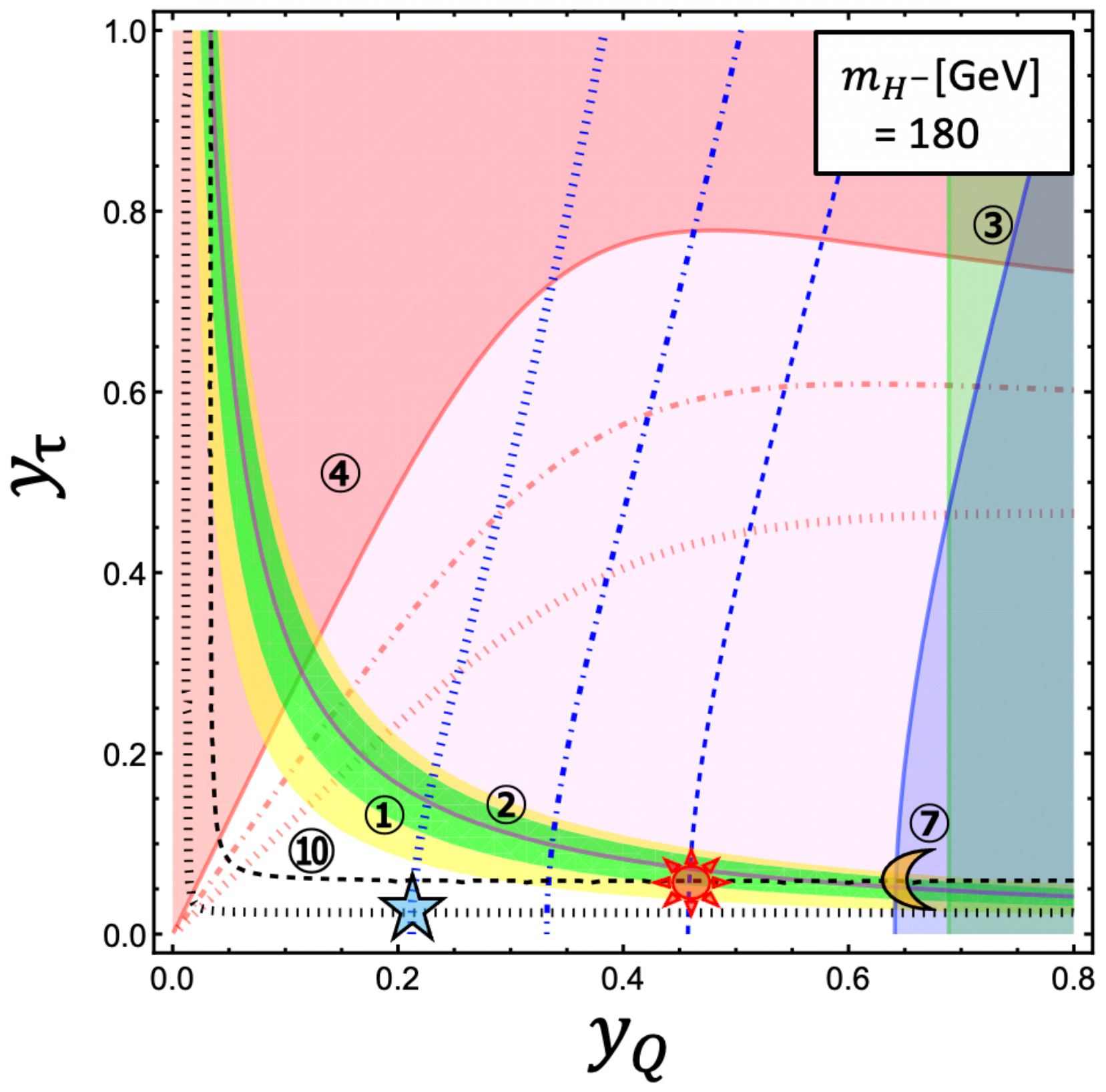}
\caption{\label{fig1}Constraints on the charged Higgs couplings $y_Q$ and $y_\tau$, for $m_{H^-} = 180\,\text{GeV}$. The green and yellow bands show the $1\sigma$ and $2\sigma$ regions implied by $R(D^{(*)})$ data, while the dashed black contour depicts the reach of the proposed $b\tau\nu$ search. See the original publication \cite{Blanke:2022pjy} for further details.}
\end{figure} 
Note that the Yukawa couplings $y_Q$, $y_\tau$ are in general complex.
Already with the currently available data, the signature $pp\to b\tau\nu$ is capable of probing a large fraction of the light charged Higgs parameter space, as shown in Fig.\ \ref{fig1} for $m_{H^-} = 180\,\text{GeV}$. Together with the constraint from (flavoured) dijet resonance searches \cite{Iguro:2022uzz}, Run 2 data can test most of the charged Higgs parameter space implied by the anomaly. Note that for heavier charged Higgs masses, the LHC search becomes even more powerful, due to the improved kinematic distinction from the SM background.

Besides the charged Higgs,  other potential NP solutions to the $R(D^{(*)})$ anomaly introduce one or several leptoquark states. While often considered ``exotic'', leptoquarks are naturally present in any theory unifying quarks and leptons, and their experimental discovery could provide hints towards a possible unification of the SM gauge symmetry. 
In fact, leptoquark models currently are the favoured solutions to the $B$ anomalies, as they provide a neat explanation of why deviations from the SM are first seen in semileptonic decay modes. In addition, leptoquarks are also less stringently constrained by complementary flavour and collider observables, compared to other NP scenarios.

Among the various leptoquark models, probably the most popular one is  the extension of the SM by an $SU(2)_L$-singlet vector leptoquark $U_1$. This simplified model has been shown to be the only single-particle scenario that can solve both the $R(D^{(*)})$ and the $b\to s\ell^+\ell^-$ anomalies \cite{Buttazzo:2017ixm}. That solution has been shown to be consistent with complementary flavour data, in particular $B_s-\bar B_s$ mixing observables and the decays $B\to K^{(*)}\nu\bar\nu$. In addition, due to its conserved $U(1)_{B-L}$ symmetry, the leptoquark does not induce proton decay. Lastly, this scenario is also appealing from a model-building perspective, as the $U_1$ leptoquark is contained in the Pati-Salam gauge group \cite{Pati:1974yy}
\begin{equation}
G_\text{PS} = SU(4)_c \times SU(2)_L \times SU(2)_R
\end{equation}
unifying quarks and leptons. The main challenge for the comstruction of a UV-complete model is then the need for a non-trivial flavour structure in the leptoquark couplings. Numerous attempts to overcome this issue have been made in the literature, see e.\,g.\ \cite{DiLuzio:2017vat,Calibbi:2017qbu,Bordone:2017bld,Greljo:2018tuh,Balaji:2018zna,Barbieri:2016las,Blanke:2018sro,Fuentes-Martin:2022xnb}.

In what follows, we stay within the simplified model framework and merely assume the minimal $U_1$ coupling scenario that is able to address the $R(D^{(*)})$ anomaly. In this case, the vector leptoquark couples to left-handed fermions only, and its coupling matrix has the structure
\begin{equation}
\lambda^{[\tau]}_{dl} = 
\left(
\begin{array}{ccc}
0 & 0 & 0  \\
0  & 0 & \lambda_{s \tau}     \\
0  & 0  & \lambda_{b \tau}  
\end{array}
\right) \,.
\end{equation}
This so-called $\tau$-isolation pattern can be motivated from residual family symmetries \cite{Bernigaud:2019bfy,Bernigaud:2020wvn}. Couplings to up-type quarks and neutrinos are implied by $SU(2)_L$ symmetry. Note that additional couplings to second-generation leptons, as required by the $b\to s\ell^+\ell^-$ anomalies, are significantly smaller than the third-generation lepton couplings, and hence the LHC phenomenology discussed below does not change qualitatively in their presence.

Interestingly, the measurements of $R(D^{(*)})$ and direct LHC searches provide complementary information on the model's parameter space \cite{Bernigaud:2021fwn}. While $R(D^{(*)})$ data determine the combination $\lambda_{s\tau}\lambda_{b\tau}/M^2$, the individual parameters can only be fixed with the help of high-$p_T$ data: The leptoquark mass $M$ can be determined through measuring the cross-section of leptoquark pair production and the invariant mass of the leptoquark decay products. The ratio of the couplings $\lambda_{b\tau}/\lambda_{s\tau}$ is then fixed by measuring the branching ratios into third- vs. second-generation final state quarks. 

\begin{figure}[t]
\centering
\includegraphics[width=0.5\textwidth]{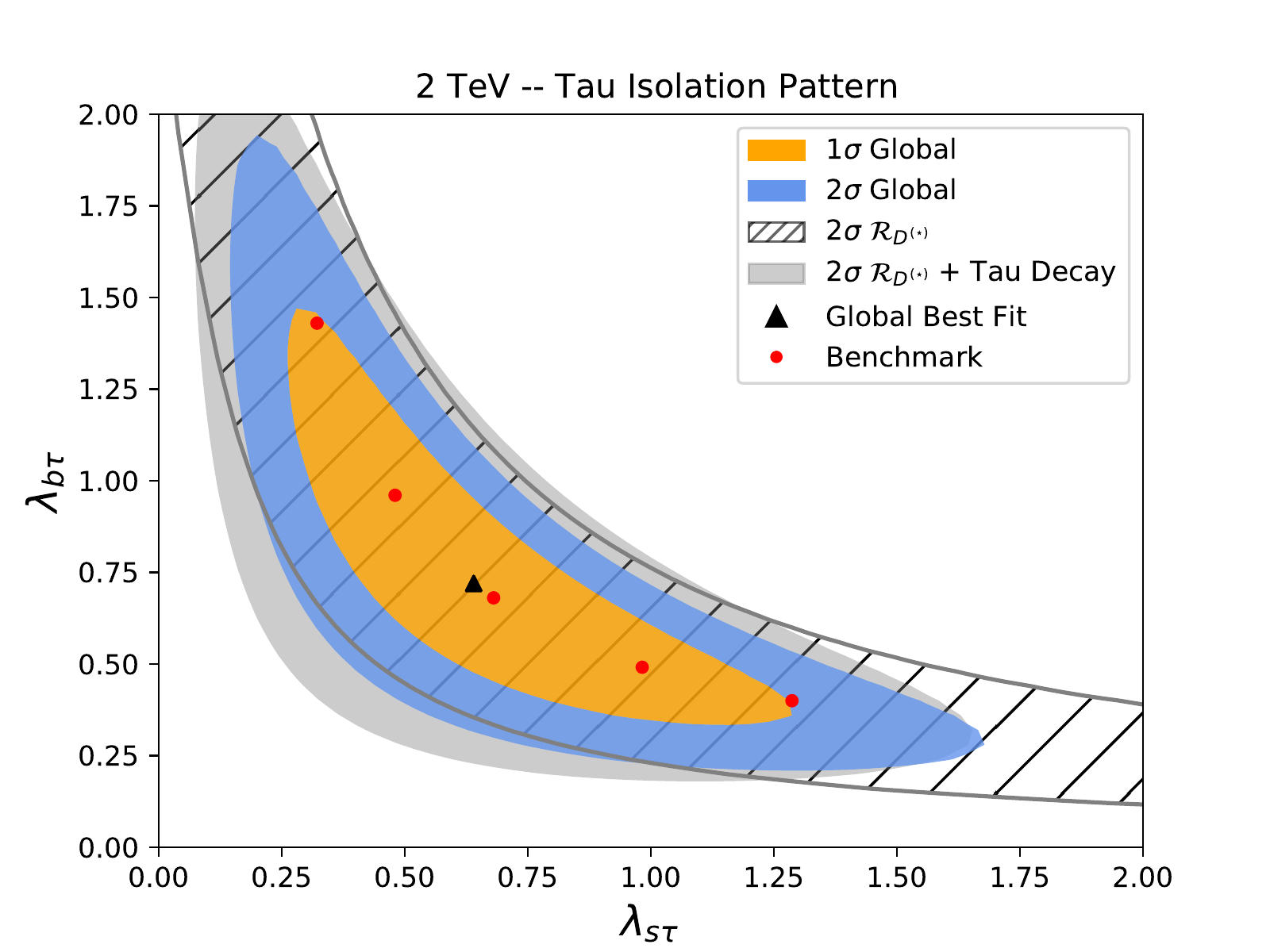}~~\includegraphics[width=0.44\textwidth]{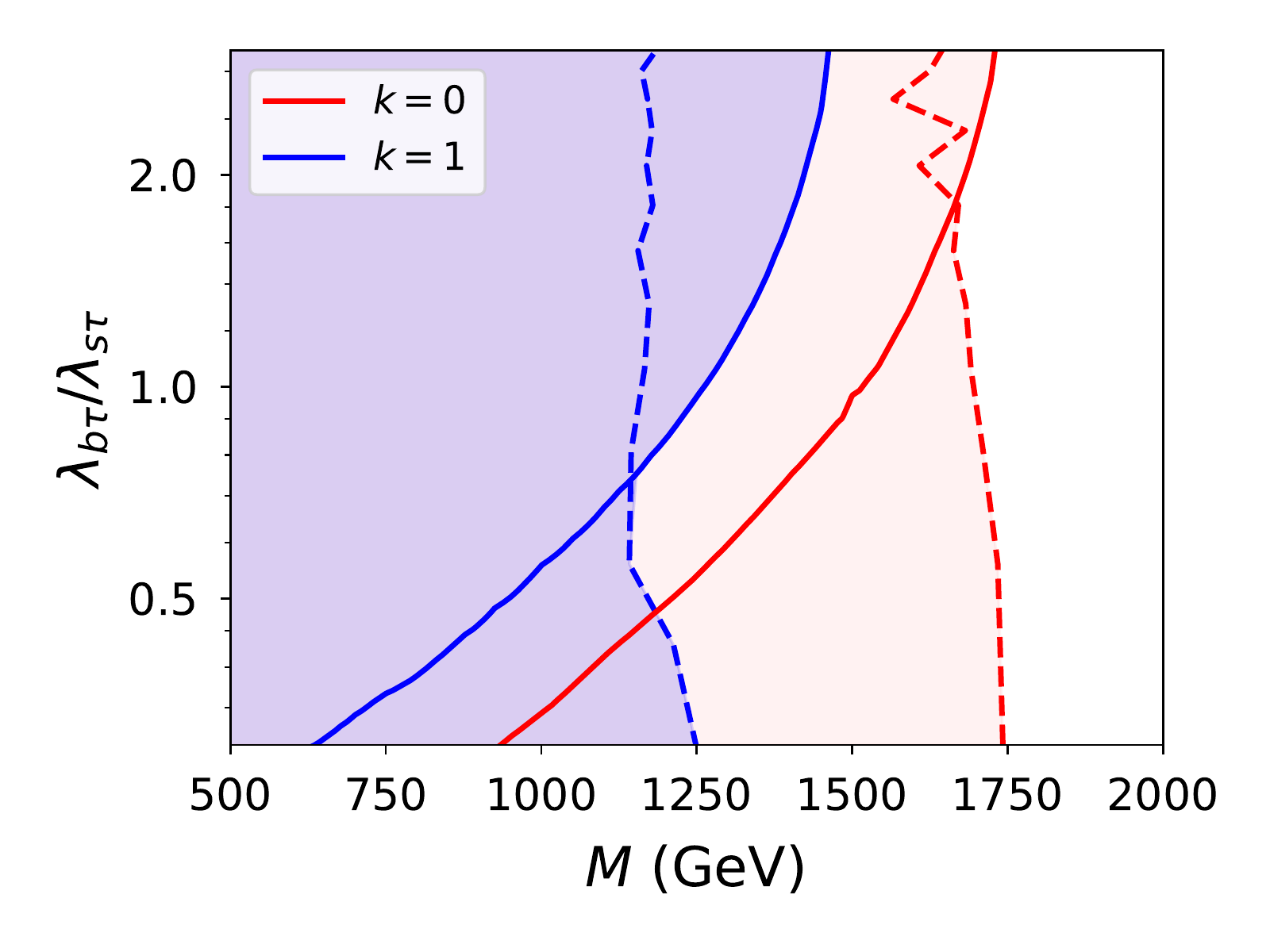}
\caption{\label{fig2}{\it Left:} Current flavour fit result on the leptoquark coupling parameters for fixed mass. The red points are benchmark point for the subsequent LHC analysis, chosen within the $1\sigma$ region.
{\it Right:} LHC exclusion contours from leptoquark pair production in the final states $b\tau\,t\nu$ (solid) and $\not\! E_T$ searches (dashed). Further details can be found in the original publication \cite{Bernigaud:2021fwn}.}
\end{figure}

Fig.\ \ref{fig2} shows the current status of this procedure, as performed in \cite{Bernigaud:2021fwn}. While the flavour data (left panel) constrain the product of coupling parameters $\lambda_{s\tau}\lambda_{b\tau}$, the red benchmark points with vastly different ratios $\lambda_{b\tau}/\lambda_{s\tau}$ lead to equal $R(D^{(*)})$ values. However, the quark-generation specific leptoquark search for the $b\tau\,t\nu$ final state 
\cite{CMS:2020wzx,ATLAS:2021jyv,Belanger:2021smw} is highly sensitive to the latter ratio, which leads to a strong dependence of the mass reach of the current experimental analysis on the latter ratio (right panel, solid contours). Contrary, searches for final states with missing transverse energy (right panel, dashed contours) are more inclusive, resulting in a much weaker dependence on the leptoquark couplings. If eventually leptoquarks are discovered in future direct searches, a full exploration of their decay channels will be crucial to fully determine the parameters of the model and thereby gain insight on the underlying flavour structure.


\section{\boldmath Anomalies in $b \to s \ell^+\ell^-$ transitions}

Another intriguing set of anomalies has emerged in observables related to $b\to s\ell^+\ell^-$ transitions. The first hint for a deviation from the SM has been found by LHCb in the angular distribution of the $B\to K^*\mu^+\mu^-$ decay \cite{LHCb:2013zuf,LHCb:2020lmf}, followed by tensions in the LFU ratios \cite{Aaij:2017vbb,Aaij:2019wad,LHCb:2021trn}
\begin{equation}
R(K^{(*)}) = \frac{\text{BR}(B\to K^{(*)}\mu^+\mu^-)}{\text{BR}(B\to K^{(*)}e^+e^-)}
\end{equation}
and some less significant discrepancies in $\text{BR}(B_s\to\mu^+\mu^-)$ \cite{ATLAS:2018cur,CMS:2019bbr,LHCb:2021awg}, $\text{BR}(B_s\to\phi\mu^+\mu^-)$ \cite{LHCb:2021zwz} etc.

On the theory side, $b\to s\ell^+\ell^-$ transitions are described by the effective Hamiltonian
\begin{equation}
\mathcal{H}_\text{eff}= -\frac{4 G_F}{\sqrt{2}} V_{tb}^* V_{ts} \frac{e^2}{16\pi^2}\sum_i(C_i  O_i +C'_i {O}'_i)+\text{h.c.}\,,
\end{equation}
where the operators most sensitive to NP are:
\begin{itemize}
\item $O_{7\gamma}^{(\prime)}=\frac{m_b}{e}(\bar s\sigma_{\mu\nu} P_{R(L)} b)F^{\mu\nu}$: This operator is always loop-induced, and it governs the $b\to s\gamma$ transition. Its contribution to $B\to K^*\ell^-\ell^-$ is enhanced in the $q^2\to 0$ limit;
\item $O_{9,10}^{(\prime)} = (\bar sb)_{V\mp A} (\bar\ell\ell)_{V,A}$: The semileptonic four-fermion operators are loop-suppressed in the SM, but can be generated at tree level in the presence of NP.
\end{itemize}

\begin{figure}[t]
\centering{\includegraphics[width=0.45\textwidth]{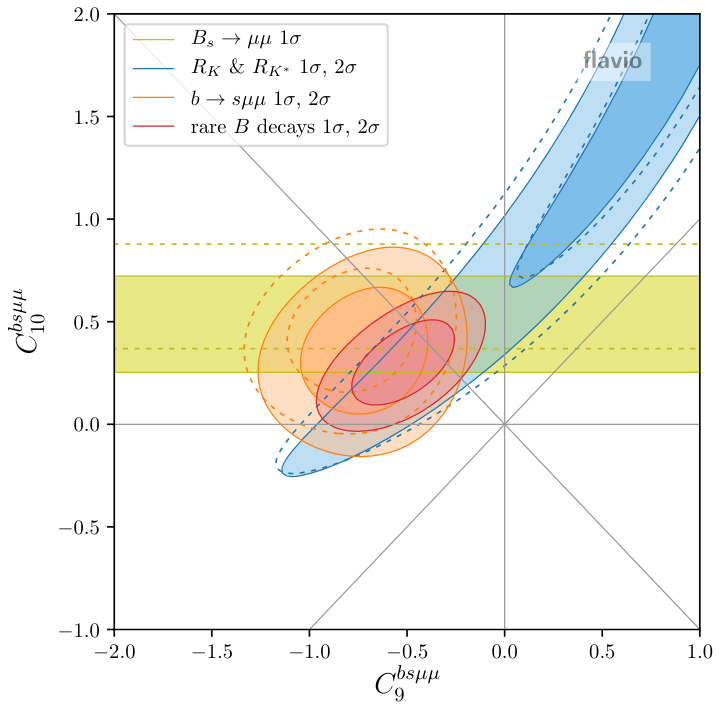}}
\caption{\label{fig3}Global fit of the Wilson coefficients $C_{9,10}^\mu$ to the $b\to s \ell^+\ell^-$ data. Figure taken from \cite{Altmannshofer:2021qrr}, where further details can be found.}
\end{figure}
Global fits of the corresponding Wilson coefficients to $b\to s\ell^+\ell^-$ data show that even the simple one-dimensional NP scenarios
\begin{equation}
\delta C_9^\mu \sim -0.73 \hspace{2cm}
\delta C_9^\mu = - \delta C_{10}^\mu \sim -0.39
\end{equation}
lead to a significant improvement of the quality of the fit, compared to the SM-only hypothesis \cite{Alguero:2021anc,Altmannshofer:2021qrr}. The latter scenario turns out to be slightly preferred due to the suppression of $\text{BR}(B_s\to\mu^+\mu^-)$ relative to the SM and due to the relative size of NP effects in $R(K)$ and $R(K^{(*)})$. It is also attractive from a model-building perspective, as it requires NP only in the interactions of left-handed fermions. The two-dimensional fit result for $\delta C_9^\mu,\delta C_{10}^\mu \ne 0$ is shown in Fig.\ \ref{fig3}. Note that a small flavour-universal NP contribution to $C_9$, possibly generated by loop-effects, further improves the goodness of fit \cite{Crivellin:2018yvo}.

Numerous NP models explaining the anomalies have been discussed in the literature. Since the required size of NP effects is substantially smaller than in the case of $R(D^{(*)})$, loop-induced NP is a viable option, see e.\,g.\ \cite{Arnan:2016cpy,Arnan:2019uhr,Belanger:2015nma,Kamenik:2017tnu}. However, also tree-level NP contributions are possible, either from $Z'$ gauge bosons \cite{Altmannshofer:2013foa,Gauld:2013qja,Altmannshofer:2014cfa,Crivellin:2015lwa,Descotes-Genon:2017ptp,DiLuzio:2019jyq,Crivellin:2022obd} or leptoquarks \cite{Hiller:2014ula,Alonso:2015sja,Fajfer:2015ycq,Calibbi:2015kma,Becirevic:2016yqi}. Note that the $bsZ'$ coupling is stringently constrained by $B_s-\bar B_s$ mixing data, requiring in turn a rather large $Z'$ coupling to muons \cite{Altmannshofer:2013foa,DiLuzio:2019jyq}. Among the various leptoquark models, the case of the $SU(2)_L$-singlet vector leptoquark sticks out, since -- as mentioned before -- it is the only single-particle solution to both the $b\to s \ell^+\ell^-$ anomalies.

The scale of NP can be estimated by translating the global fit result into the SMEFT Lagrangian, finding that the minimal solution of the $b\to s\ell^+\ell^-$ anomalies requires the operator
\begin{equation}
(\bar Q_2 Q_3)(\bar L_2 L_2)
\end{equation} 
to be generated at the scale $\Lambda_\text{NP} \sim 40\,\text{TeV}$. Here $Q_i$ and $L_i$ denote the $i$th generation left-handed quark and lepton doublet, respectively. Clearly such a large NP scale precludes the direct production of new particles at the (HL-)LHC, and searches for deviations from the SM in high-$p_T$ di-muon tails have been shown to have a limited reach \cite{Greljo:2017vvb}. 
A signifcantly larger part of the NP model space underlying the $b\to s\ell^+\ell^-$ anomalies could be probed at a future hadron collider with collision energies around $100\,\text{TeV}$, and particularly promising results are to be expected from a muon collider which should be able to probe most of the NP parameter space \cite{Huang:2021biu,Altmannshofer:2022xri,Azatov:2022itm}. It is worth noting that the latter would also allow us to explore the potential NP behind the long-standing $(g-2)_\mu$ anomaly \cite{Capdevilla:2021rwo}.


\section{Summary}

In the absence of a NP signal in direct LHC searches, the flavour anomalies are among the strongest hints for the presence of new particles at or around the TeV scale.

The $R(D^{(*)})$ anomaly, on the one hand, persists; however it is scrutinised by related flavour observables, note in particular the $R(\Lambda_c)$ sum rule. Due to the requiired low NP scale, direct searches for the underlying new particles at the LHC are very promising, and their complementarity with flavour observables will turn out useful in the complete determination of the underlying NP parameter space.

The $b\to s\ell^+\ell^-$ anomalies, on the other hand, can consistently be resolved within a global NP fit, requiring as little as one new parameter in the effective Lagrangian description. Generically, in case of new tree-level contributions, the underlying NP scale lies well beyond the reach of the LHC, requiring the construction of a new high(er)-energy hadron or muon collider. The latter endeavor is also motivated by the notorious $(g-2)_\mu$ anomaly. 

In either case it is crucial to explore
the complementarity between flavour and collider data in order to fully unravel the underlying NP and its flavour structure.


\paragraph{Acknowledgements}

I would like to thank the organisers for inviting me to present this overview at the LHCP 2022 conference.
My work is supported by the  Deutsche Forschungsgemeinschaft (DFG, German Research Foundation) under grant  396021762 -- TRR 257.

\end{document}